# The Interdecadal Bipolar Oscillation: An Atmospheric Water Vapor Mode Driving Asynchronous Polar Climate Change

Hongyu Wang[1], Jingfang Fan[1,4], Fei Xie[1], Jingyuan Li[3], Rui Shi[3], Yan Xia[1*], Deliang Chen[2*], Xiaosong Chen[1,3*]

[1] *School of Systems Science / Institute of Nonequilibrium Systems, Beijing Normal University, Beijing 10087, China.*

[2] *Department of Earth System Science, Tsinghua University, Beijing 100084, China.*

[3] *Institute for Advanced Study in Physics and School of Physics, Zhejiang University, Hangzhou 310058, China.*

[4] *Potsdam Institute for Climate Impact Research, Potsdam 14412, Germany*

* Corresponding authors:

Dr. Yan Xia (xiayan@bnu.edu.cn)

Dr. Deliang Chen (deliangchen@tsinghua.edu.cn)

Dr. Xiaosong Chen (chenxs@bnu.edu.cn)

**Abstract**

Climate change is progressing asynchronously between the Arctic and Antarctic, with important implications for global climate dynamics. While the Arctic has experienced rapid warming and pronounced amplification, the Antarctic has exhibited a delayed and heterogeneous response. Here, we identify an Interdecadal Bipolar Oscillation (IBO) in atmospheric water vapor—a coherent internal mode of variability that connects the two polar regions and helps explain their divergent climate trajectories over the past eight decades. Using reanalysis data alongside historical and pre-industrial control simulations from CMIP6, we demonstrate that the IBO is a robust internal variability mode with a quasi-period of 60–80 years. This oscillation modulates the background warming signal, with a phase shift in the late 1980s amplifying moistening and warming in the Arctic while concurrently suppressing changes in the Antarctic. Crucially, model projections suggest a possible phase reversal, which could slow water vapor increases in the Arctic while accelerating them in the Antarctic—potentially marking the start of rapid Antarctic climate change. While uncertainties remain in how climate models capture polar water vapor, our findings highlight the IBO as a pivotal driver of polar climate evolution and a potential contributor to emerging climate risks in the Southern Hemisphere.

# 1. Introduction

The polar climate system has drawn increasing attention due to its far-reaching regional and global impacts[1,2]. Rapid changes in temperature, atmospheric water vapor, precipitation, and sea ice extent are reshaping polar ecosystems[3,4], contributing to global sea-level rise[5], and influencing weather and climate patterns at mid-latitudes[6,7]. Among these factors, atmospheric water vapor plays a crucial role as both a regulator of the Earth's energy balance[8,9] and a driver of the hydrological cycle[10,11] and atmospheric circulation[12]. Given the heightened sensitivity of the Arctic and Antarctic to external forcings[13], understanding variability in polar atmospheric water vapor is especially important. However, despite its importance, this component of the polar climate remains underexamined, limiting our understanding of its variability and its broader implications for climate dynamics[14,15].

Evidence from both observations and models points to clear asymmetries in climate change between the Arctic and Antarctic[16,17]. The Arctic has experienced a steady decline in sea ice extent[18], while the Antarctic saw modest increases in earlier decades followed by a sharp decline in recent years[19]. Arctic surface temperatures have risen rapidly, exhibiting pronounced Arctic amplification, especially in winter[20], whereas Antarctic temperature trends are more spatially variable[21]—marked by cooling in East Antarctica and warming in West Antarctica and the Antarctic Peninsula[22]. Global climate models broadly capture this pattern, typically simulating weaker warming in the Antarctic compared to the Arctic in recent decades[23].

Understanding how changes at one pole relate to changes at the other remains a major scientific challenge[24–27]. Over the past century, multidecadal temperature variability at the poles has often shown a marked anticorrelation—a phenomenon commonly referred to as the bipolar seesaw[24]. While the mechanisms and persistence of this pattern remain debated[28,29], evidence from paleoclimate records, observational datasets, and model simulations supports its presence[24,30–33].

Here, we identify a robust interdecadal teleconnection in atmospheric water vapor variability between the Arctic and Antarctic over the past 82 years, based on reanalysis datasets and model outputs. We term this phenomenon the Interdecadal Bipolar Oscillation (IBO). We show that the IBO is an intrinsic internal mode of the climate system, characterized by a quasi-periodic (60–80 years) seesaw pattern between the Arctic and Antarctic. By modulating the background global warming signal, the IBO provides a physical explanation for the observed asynchronous climate trends—specifically, why Antarctic moistening stalled while the Arctic accelerated after the late 1980s. Recognizing this oscillation offers new insights into the mechanisms governing polar amplification and improves our ability to project future critical transitions in the high latitudes.

## 2. Materials and methods

### 2.1 Reanalysis

In this study, we primarily use the fifth-generation atmospheric reanalysis (ERA5),

produced by the European Centre for Medium-Range Weather Forecasts[34], which is widely recognized as a state-of-the-art reanalysis dataset, covering the period 1940–2022 with monthly temporal resolution and a horizontal resolution of 0.25° × 0.25°. We use the Total Column Water variable in ERA5 to represent atmospheric water vapor. We also use surface air temperature, sea surface temperature, evaporation, total precipitation, and the vertical integral of northward water vapor flux from ERA5 to investigate associated climate dynamics and water vapor sources.

To verify the robustness of our results, we analyze monthly mean water vapor data from the NOAA-CIRES-DOE 20th Century Reanalysis Version 3 (20CR; 1836–2015, 1° × 1°)[35] and the Modern-Era Retrospective Analysis for Research and Applications, Version 2 (MERRA-2; 1980–2022, 0.5° × 0.625°)[36]. We employ annual mean water vapor paleoclimate reanalysis data from the Last Millennium Reanalysis (LMR) Project Global Climate Reconstructions Version 2, which spans 1700–2000 with a horizontal resolution of 2° × 2°. This dataset includes 20 ensemble members, allowing us to compute the ensemble mean.

## 2.2 Climate model experiments

We utilize outputs from eight Coupled Model Intercomparison Project Phase 6 (CMIP6) models (Table S1). These models were selected from an initial pool of 49 based on their fidelity in reproducing the spatial patterns and temporal evolution of the IBO consistent with ERA5 during the 1940–2022 period. This rigorous screening ensures that our analysis relies on the subset of models that best captures the dynamics of polar water vapor variability.

We analyze pre-industrial control (piControl) simulations to investigate internal variability. For the historical period (1850–2014), in addition to standard simulations forced by all natural and anthropogenic factors , we analyze detection and attribution experiments driven individually by natural forcings, greenhouse gases, and anthropogenic aerosols. Future projections (2015–2100) are derived from four Shared Socioeconomic Pathway (SSP) scenarios (SSP1-2.6, SSP2-4.5, SSP3-7.0, and SSP5-8.5). These future projections are concatenated with the corresponding historical simulations to form continuous time series.

For all experiments, we use only the first ensemble member (r1i1p1) from each model. All model outputs were regridded to a common resolution of 1.875° × 1.25° for multi-model comparison.

**2.3 Eigen microstate theory**

The eigen microstate theory (EMT) is a universal theoretical framework for the analysis of complex systems based on Gibbs' ensemble theory in statistical physics. EMT has demonstrated its potential in revealing intrinsic structures and critical behaviors of complex Earth systems, as shown in studies on surface air temperature[37,38], stratospheric ozone[39,40], and sea surface temperature in the western Pacific [41].

Here we consider the global water vapor in the atmosphere as a complex system consisting of $N$ grid points as agents with $M$ measurements. Given the significant spatial variability in water vapor magnitude, a microstate at time $t$ is represented as a vector $\boldsymbol{S}(t) = [s_1(t), s_2(t), ..., s_N(t)]^T$, where $s_i(t)$ denotes the water vapor anomaly at grid $i$ standardized by its standard deviation $\Delta_i$. By collecting $M$ samples, we construct an $N \times M$ ensemble matrix $\boldsymbol{A}$ with entries $A_{it} = s_i(t)/\sqrt{C_0}$, where the normalization factor is given by $C_0 = \sum_{t=1}^{M} \sum_{i=1}^{N} s_i^2(t)$.

The eigen microstates $\{\boldsymbol{u}_k\}$, their temporal evolutions $\{\boldsymbol{v}_k\}$, and the corresponding eigenvalues $\lambda_k$ are defined by a singular value decomposition as $\boldsymbol{A} = \sum_k \lambda_k \boldsymbol{u}_k \boldsymbol{v}_k^T$. Here, $\lambda_k^2$ quantifies the probability distribution (weight) of these eigen microstates. The eigen microstates represent recurrent, system-wide patterns of co-varying water vapor anomalies. Specifically, $\boldsymbol{u}_k$ reveals the spatial structure, while $\boldsymbol{v}_k$ reflects the intensity and direction of the evolution over time.

We can reconstruct the eigen evolution of the $k$-th eigen microstate which is calculated as $\boldsymbol{W}_k^{rec} = \sqrt{C_0} \cdot \boldsymbol{\Delta} \odot (\lambda_k \boldsymbol{u}_k \boldsymbol{v}_k^T)$, where $\odot$ denotes the Hadamard product. Consequently, the observed water vapor anomaly is the linear superposition of all eigen evolutions. Therefore, reconstruction based on eigen microstates enables us to quantitatively separate the contribution of a specific eigen microstate.

We provide a more detailed explanation of EMT, highlighting its theoretical foundations, methodological innovations, and key differences from the commonly used empirical orthogonal function[42] (Supplementary Information, section 1). To address concerns that the high density of grid points at high latitudes might artificially amplify polar variability, we validate our findings using polar-only eigen microstates, IBO index-based projections, and area-weighting sensitivity tests (Supplementary Information, section 2). These efforts confirm that the IBO signal is robust, physically meaningful, and not an artifact of methodological choices.

**2.4 Climate indices**

Arctic warming amplification is widely defined as the ratio of linear temperature trends between the Arctic and the global mean temperature[43–47]. While boundary definitions vary[44,45,48], we define the Arctic and Antarctic as the regions poleward of 60°N and 60°S, respectively. Accordingly, the Polar Warming Amplification Index (PWAI) is calculated as the ratio of polar to global temperature trends. Similarly, we introduce the Polar Moistening Amplification Index (PMAI), defined as the ratio of linear trends in the percentage change of water vapor relative to climatology between the polar and global means. This percentage-based approach accounts for large regional variations in base water vapor content. These indices are computed using a 40-year sliding window to capture multidecadal variability (Supplementary Information, section 3). An index greater than 1 indicates amplification, while a value between 0 and 1 (or negative) implies polar changes are slower than or opposite to global trends.

We define the IBO index as $(TCW_N - TCW_S)/2$ where the $TCW_N$ or $TCW_S$ is the water vapor anomaly averaged over the Arctic and Antarctic, respectively. The IBO index can be projected onto the ensemble matrix $A$ to derive its corresponding spatial pattern. The Atlantic Multidecadal Oscillation (AMO) index is calculated from the relevant datasets for comparison with the IBO derived from different sources. The AMO index is defined as a detrended, weighted average of sea surface temperature (SST) over the 0° to 65° N region of the North Atlantic Ocean. El Niño–Southern Oscillation (ENSO) events are characterized by the Oceanic Niño Index (ONI), calculated as the 3-month moving average of ERSST.v5 SST anomalies in the Niño 3.4 region (5°S–5°N, 170°W–120°W), with a threshold of ±0.5°C for at least five consecutive months.

**2.5 Statistical analysis**

Prior to analysis, all monthly data are deseasonalized by removing the annual cycle. We verify that the leading modes are independent of the deseasonalized method (Fig. S9). To focus on interdecadal variability, a 16-year moving average is applied to the time series. Similar results are obtained using moving averages ranging from 10 to 20 years (Fig. S10). The statistical significance of correlation coefficients and linear trends was assessed using the Student's t-test.

## 3. Results

### 3.1 Interdecadal variability in Arctic and Antarctic atmospheric water vapor

Amid ongoing global warming, observations indicate a steady rise in atmospheric water vapor over recent decades[49,50]. Globally, surface air temperature has increased by 0.13 °C per decade, accompanied by a rise in atmospheric water vapor of 0.15 kg/m² per decade over the past 83 years (Fig. S11). However, the polar regions exhibit distinct interdecadal variability that diverges from global trends, particularly through contrasting patterns between the Arctic and Antarctic (Fig. 1a, b). Analysis of these changes reveals two distinct phases between 1940 and 2022, separated by a critical turning point around 1987. This transition marks a shift in polar climate regimes and highlights the growing divergence in polar responses, especially from the late 1980s onward.

During the first phase (1940–1987), the Arctic experienced only modest warming alongside a slight decline in water vapor (0.05 °C per decade, −0.02 kg/m² per decade). In contrast, the Antarctic showed stronger increases in both surface air temperature and water vapor (0.17 °C per decade, 0.07 kg/m² per decade). The second phase (1988–2022) reveals a reversal: the Arctic underwent rapid increases in both surface air temperature and water vapor (0.68 °C per decade, 0.21 kg/m² per decade), while the Antarctic recorded only weak increases (0.09 °C per decade, 0.01 kg/m² per decade).

To assess the magnitude of polar changes relative to global averages, we applied a polar amplification index to both warming and moistening (Fig. 1c, d). Results show that amplification was stronger in the Antarctic before 1987, but has since become much more pronounced in the Arctic. During 1940–1987, the moistening amplification index was −0.39 for the Arctic and 2.38 for the Antarctic. In the 1988–2022 period, the indices shifted to 2.38 for the Arctic and 0.18 for the Antarctic.

These contrasting trends in surface air temperature and atmospheric water vapor—both in absolute terms and relative to global averages—highlight the asymmetric nature of climate change at the two poles. This raises a central question: what drives the divergent water vapor changes in the Arctic and Antarctic under the shared influence of global warming?

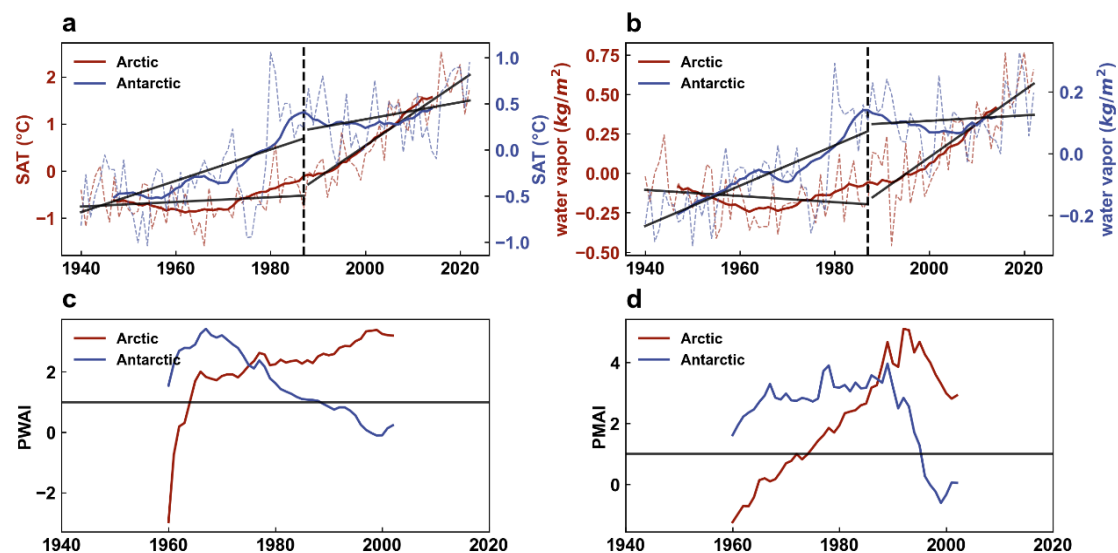

**Fig. 1. Asymmetric climate change signals between the Arctic and Antarctic. a, b,** Time series of area-weighted surface air temperature (SAT) anomalies (**a**) and water vapor anomalies (**b**) for the Arctic (red) and Antarctic (blue), shown as annual means (dashed lines) and 16-year moving averages (solid lines). Black solid lines indicate linear trends for distinct time periods, and the black dashed vertical line marks 1987, the onset of the Antarctic trend shift. **c, d,** Time series of the Arctic (red) and Antarctic (blue) polar warming amplification index (PWAI; **c**) and polar moistening amplification index (PMAI; **d**), computed using a 40-year sliding window. The black horizontal line at 1 denotes the threshold for polar amplification relative to the global mean.

To explore this, we used EMT to isolate polar-scale variability while minimizing the dominant influence of tropical variability in global water vapor fields. Results reveals that polar water vapor anomalies are primarily governed by two principal modes (Fig. 2). These modes are also evident in both NOAA-CIRES-DOE 20th Century Reanalysis (20CR; 1836–2015) and the Modern-Era Retrospective Analysis for Research and Applications, Version 2 (MERRA-2; 1980–2022) datasets (Fig. S12a–h).

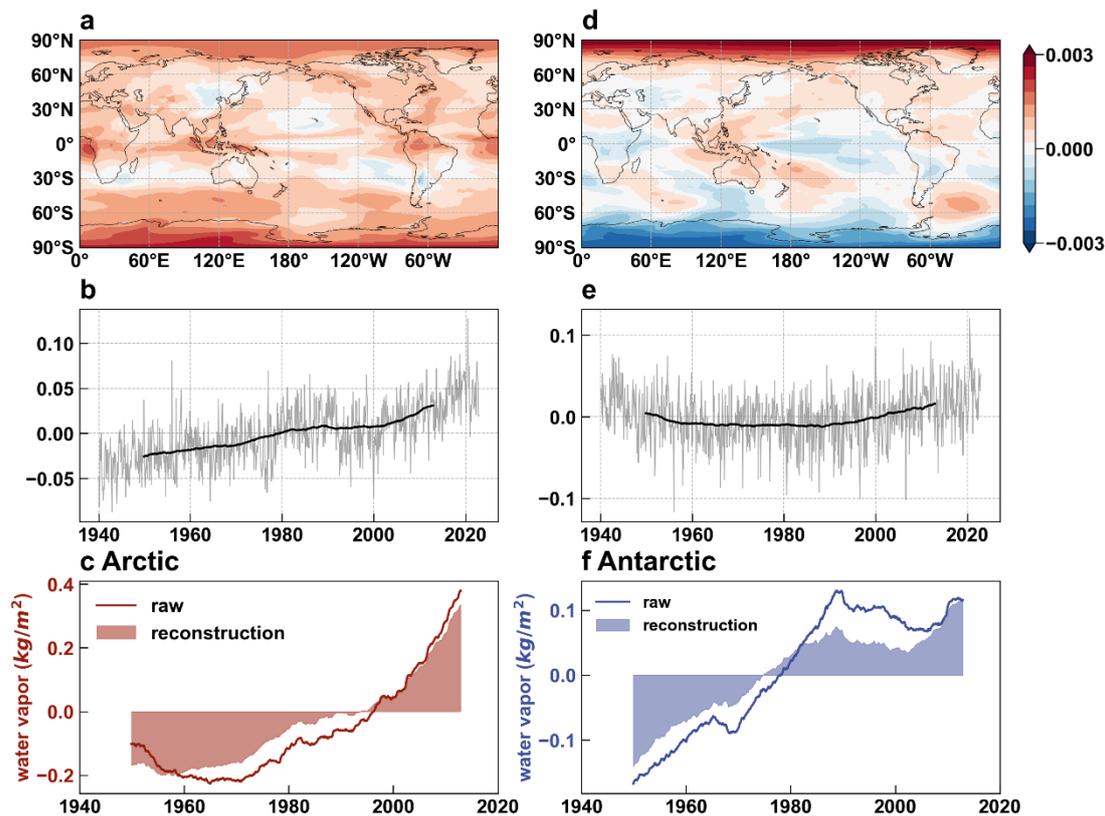

**Fig. 2. The Interdecadal Bipolar Oscillation (IBO) in atmospheric water vapor.** Spatial patterns and temporal evolution of the two leading modes of polar water vapor variability, and their contributions to reconstructed polar anomalies. **a, b,** Spatial pattern (**a**) and corresponding temporal evolution (**b**) of the first leading mode, representing the synchronous-change mode. **d, e,** Spatial pattern (**d**) and corresponding temporal evolution (**e**) of the second leading mode, identified as the IBO. Grey lines show monthly anomalies and black lines indicate 16-year moving averages. **c, f,** Time series

of 16-year moving average water vapor anomalies in the Arctic (**c**, red) and Antarctic (**f**, blue). Solid lines represent observed anomalies, and shaded areas indicate reconstructed values based on the two leading modes.

The first mode displays a spatial pattern of widespread water vapor increases, including in both polar regions (Fig. 2a), with a steadily rising temporal evolution over time (Fig. 2b). Despite relatively low absolute moisture levels at the poles, standardized variability is greatest in these regions. This mode, termed the "synchronous-change mode", is strongly correlated with global temperature increases ($r = 0.91$, $p < 0.05$; Fig. S13), reflecting the thermodynamic expectation that warmer air holds more moisture[51].

The second mode exhibits a clear out-of-phase spatial pattern between the Arctic and Antarctic (Fig. 2d), with temporal evolution showing a decline before the 1980s and an increase thereafter (Fig. 2e). This suggests that increases in Antarctic water vapor before 1987 coincided with decreases in the Arctic, and vice versa after 1987. We refer to this mode as the IBO, which captures a teleconnected pattern of polar water vapor variability. We define the IBO index as the difference in water vapor anomalies between the Arctic and Antarctic, and we derive its spatial pattern through projection (Fig. S2). The IBO index closely matches both the temporal ($r = 0.98$, $p < 0.05$) and spatial ($r = 0.90$, $p < 0.05$) characteristics of the IBO identified via the eigen microstate analysis, demonstrating its suitability as a robust and effective indicator for further research.

Reconstruction of polar water vapor anomalies using these two modes reproduces the observed variability with high accuracy (r = 0.97 and 0.96; p < 0.01; Fig. 2c, f). These modes have together dominated the large-scale, interdecadal variability in polar water vapor. Before 1987, the synchronous-change mode and the IBO reinforced each other in the Antarctic—accounting for approximately 70% and 25% of the observed moistening, respectively—while effectively cancelling out in the Arctic. Conversely, after 1987, the IBO phase shift reversed this dynamic, amplifying the Arctic signal (contributing ~35% alongside the synchronous mode's ~50%) while suppressing Antarctic moistening. These results suggest that while the synchronous-change mode captures a globally coherent moistening signal driven by global warming, the IBO adds a contrasting, oscillatory component that operates out of phase between the two poles. Understanding the dynamics and evolution of the IBO is therefore essential for anticipating future trends in polar water vapor and for improving projections of polar climate change.

**3.2 Robustness and periodicity of the IBO**

The temporal evolution of the IBO in ERA5 shows a clear long-term decline prior to the 1980s, followed by a steady upward trend, hinting at a low-frequency oscillation with a cycle of roughly 80 years. However, the relatively limited duration of the ERA5 dataset constrains our ability to robustly assess the stability and persistence of the IBO period. To address this limitation, we extended our analysis using a paleoclimate reanalysis spanning 1700–2000 and historical simulations from CMIP6, covering the

period 1850–2014.

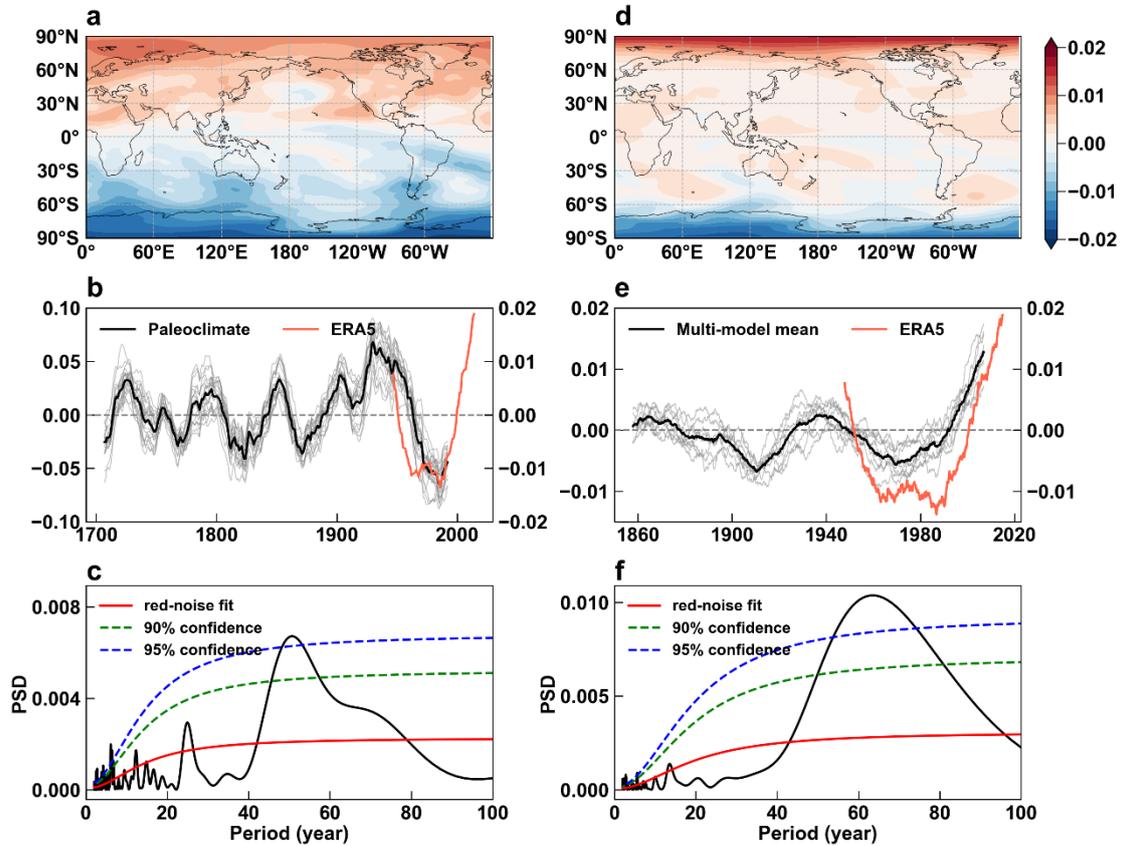

**Fig. 3. The Interdecadal Bipolar Oscillation (IBO) from paleoclimate reanalysis and CMIP6 historical simulations. a, b,** Ensemble mean spatial pattern (**a**) and 16-year moving average temporal evolution (**b**, black) of the IBO derived from paleoclimate reanalysis. Grey lines show individual ensemble members and the orange line shows the IBO from ERA5 (from Fig. 2e) for comparison. **c,** Power spectrum of the annual ensemble-mean IBO time series from paleoclimate reanalysis. The red line represents the red-noise background; green and blue lines mark the 90% and 95% significance levels, respectively. **d-f,** As in **a-c**, but based on historical simulations from eight CMIP6 models.

Out-of-phase variability between the Arctic and Antarctic emerges clearly in the spatial patterns derived from both the ensemble mean of the paleoclimate reanalysis and the CMIP6 multi-model mean (Fig. 3a, d), confirming the robustness of the IBO signal. Crucially, despite minor regional differences in lower latitudes, the antiphase variability between the Arctic and Antarctic remains highly consistent with the pattern identified in ERA5 (Fig. 2d). The corresponding temporal evolutions also exhibit pronounced interdecadal oscillations that align well with the phase shift seen in ERA5 during the overlapping period (Fig. 3b, e).

To further investigate the periodicity of the IBO, we analyzed the power spectral density of its temporal evolution (Fig. 3c, f). Results from the paleoclimate reanalysis suggest a quasi-period of 50–70 years, while CMIP6 simulations indicate a slightly longer cycle of 60–80 years. Collectively, these findings provide compelling evidence that the IBO is a persistent interdecadal oscillation in polar water vapor, with a robust quasi-period of approximately 60–80 years.

To definitively rule out the possibility that the IBO is solely a response to external forcing, we examined piControl simulations from CMIP6, in which external forcings are held constant. As illustrated by two representative models (ACCESS-CM2 and CESM2) in Fig. 4 and further confirmed across the broader models in Fig. S14, seven out of eight models successfully reproduce the bipolar spatial pattern and the characteristic 60–80-year quasi-period of the IBO. The persistence of the IBO in the absence of time-varying external forcing serves as decisive evidence that this oscillation is an inherent internal mode of the polar climate system, rather than a forced response.

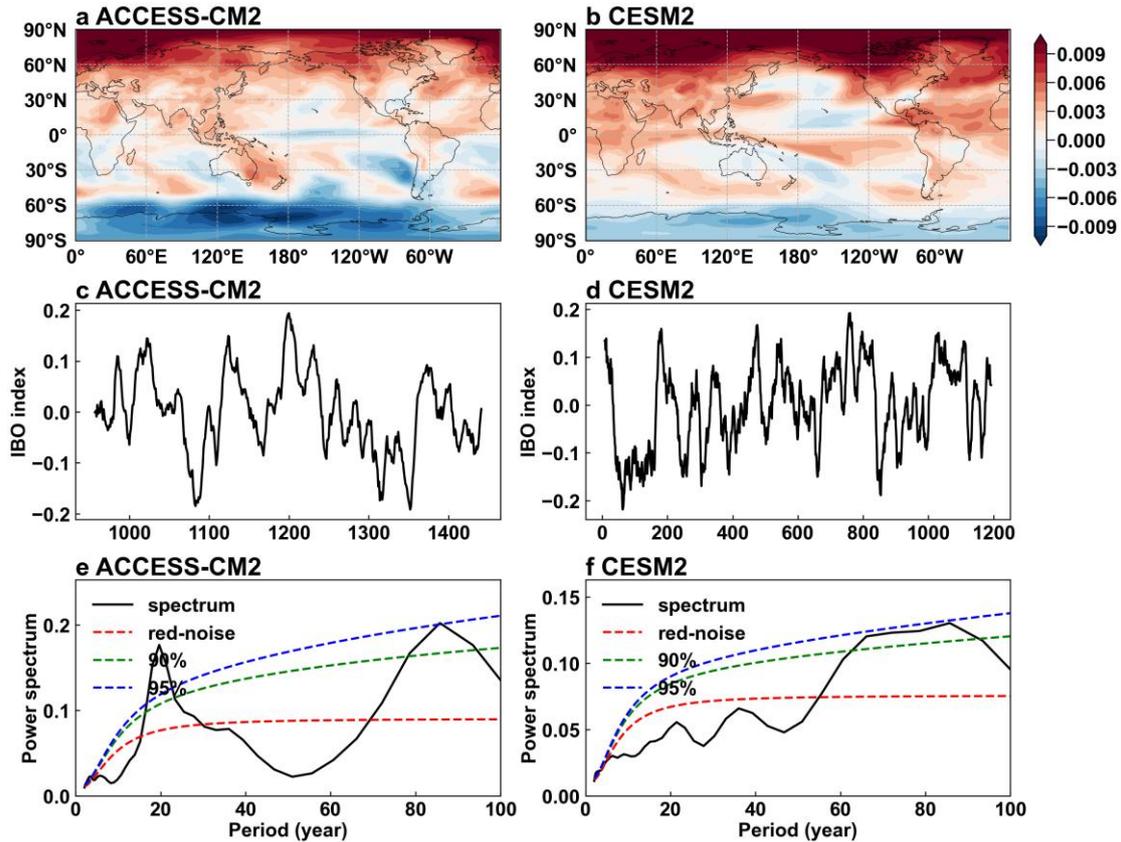

**Fig. 4. Interdecadal Bipolar Oscillation (IBO) from pre-industrial control simulations of CMIP6 models and corresponding power spectra. a, b,** Spatial patterns of the IBO for ACCESS-CM2 (**a**) and CESM2 (**b**). **c, d,** Temporal evolution of the IBO index for ACCESS-CM2 (**c**) and CESM2 (**d**), smoothed with a 16-year moving average. **e, f,** Power spectra of the IBO index for ACCESS-CM2 (**e**) and CESM2 (**f**). The red line indicates the red-noise background, and green and blue dashed lines mark the 90% and 95% significance levels, respectively. Corresponding results for all analyzed models are provided in Fig. S14.

**3.3 Hydrological origins and external modulation of IBO periodicity**

Having established the IBO as an intrinsic internal mode, we next investigate the hydrological sources driving its 60–80-year quasi-period and how its evolution is modulated by external forcing. We first examined the sources of polar water vapor, which can be broadly divided into two components: the local balance between evaporation and precipitation (E−P), and the moisture transport from lower latitudes (Fig. 5a, b).

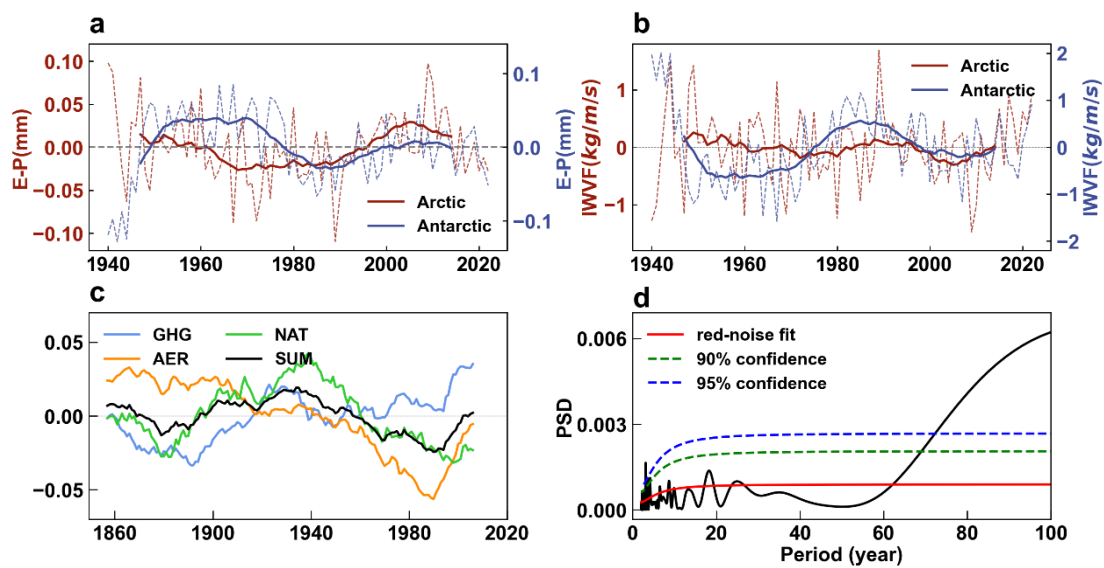

**Fig. 5. Contributions of water vapor sources and external forcings to the 60–80-year quasi-periodicity of the Interdecadal Bipolar Oscillation (IBO). a,** Time series of area-weighted, detrended evaporation minus precipitation (E−P) in the Arctic (red) and Antarctic (blue), shown as annual values (dashed lines) and 16-year moving averages (solid lines). Positive values contribute to increased water vapor within the polar regions. **b,** Same as in **a**, but for zonal mean detrended vertically integrated water vapor flux (IWVF) at 60°N (Arctic boundary, red) and 60°S (Antarctic boundary, blue). Positive values indicate an increase in water vapor within the polar regions. **c,** Reconstructed annual water vapor anomalies associated with the IBO under greenhouse gas–only forcing (GHG, blue), aerosol–only forcing (AER, orange), and natural forcing (NAT, green), based on the multi-model mean from CMIP6. The black line (SUM) represents the mean of the three individual forcing time series. **d,** Power spectrum of the annual SUM time series. The red line indicates the red-noise background; green and blue lines mark the 90% and 95% significance levels, respectively.

In the Arctic, the local E−P balance exhibits a pronounced 60–80-year oscillation that aligns closely with the IBO, whereas moisture transport shows no significant long-term variability (Fig. S15a, c). This suggests that the Arctic IBO signal is primarily driven by local hydrological dynamics. In contrast, the Antarctic IBO signal appears to emerge from a combination of processes, as both the local E−P balance and moisture transport display quasi-periodic signals of about 40–60 years (Fig. S15b, d).

These differences in water vapor sources between the two poles may stem from their contrasting geography and topography. The Arctic, characterized by ocean-dominated, low-lying terrain, is more responsive to changes in local evaporation and precipitation[52]. In contrast, the elevated Antarctic continent relies more heavily on atmospheric moisture transport from lower latitudes due to its limited local moisture sources[53].

While internally generated, the strength, phase, and periodicity of the IBO are sensitive to external modulation. We systematically assess the roles of anthropogenic and natural forcings using CMIP6 single-forcing experiments. Notably, all model ensemble members consistently reproduce the IBO across three forcing scenarios, confirming its persistence even under perturbed climate conditions (Fig. S16). Assuming linear additivity among individual forcing effects and neglecting nonlinear interactions, we reconstruct the IBO's temporal evolution by summing the water vapor anomalies associated with the IBO under three separate forcing scenarios (Fig. 5c). The resulting combined-forcing IBO reproduces an 80-year quasi-period (Fig. 5d), closely matching the cycles identified in both reanalysis data and CMIP6 historical simulations, while also maintaining a consistent phase.

Among the individual forcings, the IBO under greenhouse gas forcing sustains the 60–80-year quasi-period, along with superimposed short-term variability (Fig. S15e).

The IBO under natural forcing also shows a dominant quasi-period of near 80 years (Fig. S15g). In contrast, while the IBO under aerosol-only forcing lacks a well-defined long-term periodicity (Fig. S15f), its phase evolution aligns closely with the observed IBO shift after the 1940s (Fig. 5c). This suggests that anthropogenic aerosols may have played a role in shifting the IBO's phase around 1980, potentially linked to the sharp decline in North American and European aerosol emissions following air quality regulations in the 1980s–1990s[54].

### 3.4 The IBO and future polar moistening under climate change

Our results demonstrate that historical changes in polar water vapor are shaped by two dominant modes: the globally coherent synchronous-change mode and the IBO, whose long-term periodicity arises from internal hydrological processes but is modulated by greenhouse gas forcing. A critical question is how the projected phase evolution of the IBO will influence future Arctic and Antarctic water vapor variability under different emission pathways.

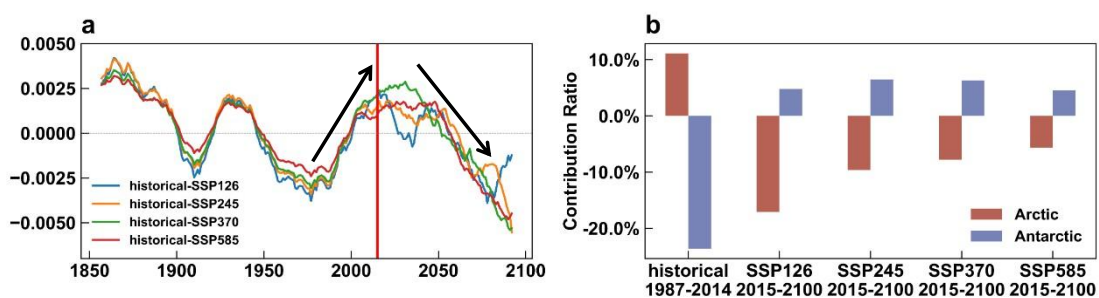

**Fig. 6. Future evolution of the Interdecadal Bipolar Oscillation (IBO) and its influence on polar moistening amplification**. **a,** Temporal evolution of the IBO (16-year moving average) under four SSP scenarios: SSP1-2.6 (blue), SSP2-4.5 (orange),

SSP3-7.0 (green), and SSP5-8.5 (red), based on the multi-model mean of CMIP6 simulations. Historical runs are concatenated with each SSP scenario to generate continuous time series. The red vertical line marks the year 2014 and black arrows indicate the trends before and after this year. **b,** Contribution of the IBO to moistening amplification in the Arctic (red) and Antarctic (blue) during the recent historical period (1987–2014) and the future projection period (2015–2100).

To assess this, we extended the CMIP6 historical simulations by incorporating projections from various SSP scenarios (Fig. S17 and Fig. 6a). Across all SSPs, the IBO is projected to transition from its recent positive phase toward a negative one in the coming decades, suggesting a reversal in its influence. Specifically, this shift implies that the IBO will change from amplifying Arctic moistening to suppressing it, while simultaneously shifting from dampening Antarctic moistening to enhancing it (Fig. 6b).

In the Arctic, the IBO's projected future negative phase may partially offset the moistening associated with the synchronous-change mode, potentially slowing the rate of Arctic moistening amplification. This trend aligns with recent observations of a slowdown in Arctic warming and stabilization in sea ice loss[55], and may also be linked to phase transitions in other decadal climate modes such as the Interdecadal Pacific Oscillation and the Arctic Mode[48]. However, if the IBO's dampening effect is insufficient, continued greenhouse gas–induced warming may still drive enhanced moistening in the Arctic.

In contrast, the implications for the Antarctic are more concerning. The IBO's shift toward enhancing Antarctic moistening is expected to reinforce the synchronous-change signal, accelerating the moistening rate in the Antarctic. This constructive interference could contribute to a more pronounced Antarctic moistening amplification, potentially leading to a scenario where the Antarctic begins to exhibit rapid amplification similar to that previously seen in the Arctic. Given that increased water vapor strengthens downward longwave radiation, these changes align with emerging evidence of accelerating Antarctic climate change, including faster warming and sea ice loss[56–59]. The IBO's projected phase reversal may thus mark the onset of a more rapid climatic shift in the Antarctic, increasing the risk of destabilizing the Antarctic ice sheet system.

## 4. Discussion and Conclusion

This study reveals interdecadal asymmetries in polar water vapor change: prior to the 1980s, the Antarctic experienced more rapid moistening, while after the 1980s, the Arctic became dominant. We identify two key modes that explain this variability: the synchronous-change mode, which tracks global warming, and the IBO, which follows a quasi-periodic bipolar seesaw pattern with a cycle of approximately 60–80 years. The IBO index developed here proves effective in capturing both the temporal evolution and spatial structure of this mode. We validate the IBO's robustness through reanalysis data, multiple fully coupled model simulations in the pre-industrial and historical periods, establishing it as a mode of intrinsic internal variability.

While we have identified the hydrological origins and forcing-induced modulations of the IBO periodicity, the detailed physical mechanisms behind it warrant further investigation. Although the IBO shows antiphase behavior reminiscent of ENSO, our analysis excluding ENSO events (Fig. S18) suggests ENSO is not its primary driver. We note that the IBO shows potential phase alignment with the Atlantic Multidecadal Oscillation (AMO) in both observations and models (Fig. S14 and Fig. S19). The AMO is known to influence Arctic climate via enhanced atmospheric rivers, increased poleward energy transport, and sea ice loss during positive phases[60–62]. At the same time, AMO-driven Rossby wave trains can alter Antarctic atmospheric circulation, affecting regional temperatures, sea ice, ocean heat uptake, and ice sheet mass balance[63]. Negative AMO phases have been associated with accelerated Antarctic warming[64]. These opposing polar effects suggest that the AMO may play a key role in generating the bipolar structure of the IBO. Trans-equatorial processes including AMO-mediated oceanic pathways and interhemispheric atmospheric teleconnections[65,66] merit further exploration.

We acknowledge systematic biases in CMIP6 models regarding polar climate processes, stemming mainly from coarse spatial resolution and inadequate parameterization of polar-specific physical mechanisms[67–69]. The models exhibit systematic biases in representing key processes such as cloud–radiation interactions and precipitation-related radiative effects, which are crucial for simulating polar amplification[70,71]. Additionally, current models struggle to accurately represent complex ice–ocean–atmosphere coupling, especially involving Antarctic topography and deep-water formation[72]. To mitigate these uncertainties, we implemented a rigorous selection process, retaining only the models that demonstrate high fidelity in reproducing the observed spatial patterns and temporal evolution of polar water vapor (as detailed in Methods). Consequently, our findings rely on the simulations most consistent with reanalysis data, thereby minimizing uncertainty.

Understanding the IBO's influence on polar moistening is critical for improving decadal-scale predictions. Under all future emission scenarios, the IBO is projected to enter a negative phase—dampening Arctic moistening and enhancing Antarctic moistening. Because increased water vapor strengthens downward longwave radiation and intensifies surface warming, these changes may accelerate ice loss and amplify regional climate feedbacks. The anticipated Antarctic moistening amplification could signal a shift toward faster climate change in the region, similar to recent rapid changes in the Arctic. These findings underscore the importance of monitoring Antarctic climate trajectories and assessing the potential for tipping points in a warming world.

**Conflict of interest**

The authors declare that they have no conflict of interest.

**Acknowledgments**

This work is supported by the National Key R&D Program of China (grant 2023YFE0109000), and the National Natural Science Foundation of China (grants 12135003 and 42105016). This work is also supported by the ClimTip project (ClimTip contribution #), which has received funding from the European Union's Horizon Europe research and innovation program under grant agreement no. 101137601. As an Associated Partner, BNU has received funding from the Chinese Ministry for Science and Technology (MOST). DC was supported by Tsinghua University (100008001). The contents of this publication are the sole responsibility of the authors and do not necessarily reflect the opinions of the European Union or MOST. We acknowledge the World Climate Research Programme, which, through its Working Group on Coupled Modelling, coordinated and promoted CMIP6. We thank the climate modeling groups for producing and making available their model output, the Earth System Grid Federation (ESGF) for archiving the data and providing access, and the multiple funding agencies who support CMIP6 and ESGF.

**Author Contributions**

H.W., Y.X., D.C and X.C. designed the research, conceived the study, carried out the analysis and drafted the manuscript. H.W., J.F., F.X., R.S., J.L., Y.X., D.C. and X.C. generated research ideas, discussed results, provided comments and revised the manuscript. All authors reviewed the manuscript.

**Data availability**

The CMIP6 outputs are available from the Earth System Grid Federation (ESGF) portal at https://aims2.llnl.gov/search. The ERA5 reanalysis data are available from the Copernicus data store at https://cds.climate.copernicus.eu/cdsapp#!/dataset/reanalysis-era5-single-levels-monthly-means?tab=form. The 20CR reanalysis data are available at https://psl.noaa.gov/data/gridded/data.20thC_ReanV3.html. The MERRA-2 reanalysis data are available at https://disc.gsfc.nasa.gov/datasets/M2IMNXASM_5.12.4/summary?keywords=M2IMNXASM_5.12.4. The paleoclimate reanalysis data from Last Millennium Reanalysis (LMR) Project Global Climate Reconstructions Version 2 are available at https://www.ncei.noaa.gov/access/paleo-search/study/27850. The ONI index is available at https://origin.cpc.ncep.noaa.gov/products/analysis_monitoring/ensostuff/ONI_v5.php.

**Supplementary information**

Supplementary information is available for this paper.